\documentclass[conference]{IEEEtran}
\usepackage{amssymb}
\usepackage{verbatim}
\usepackage{graphicx,epsfig}
\usepackage{amsfonts}
\usepackage{subfigure}
\usepackage[normalem]{ulem}
\usepackage{amsmath}
\usepackage{authblk}

\def\bi{\begin{itemize}}
\def\ei{\end{itemize}}
\def\bequ{\begin{equation}}
\def\eequ{\end{equation}}

\def\benum{\begin{enumerate}}
\def\eenum{\end{enumerate}}

\begin{document}

\title{Maximum Multipath Routing Throughput in Multirate Wireless Mesh Networks}

\author[1]{Jalaluddin Qureshi}
\author[2]{Chuan Heng Foh}
\author[3]{Jianfei Cai}

\affil[1]{Department of Electrical Engineering, Namal College, Mianwali, Pakistan}

\affil[2]{CCSR, Department of Electronic Engineering, University of Surrey, UK}

\affil[3]{School of Computer Engineering, Nanyang Technological University, Singapore}

\maketitle

\begin{abstract}
In this paper, we consider the problem of finding the maximum routing throughput between any pair of nodes in an arbitrary multirate wireless mesh network (WMN) using multiple paths. Multipath routing is an efficient technique to maximize routing throughput in WMN, however maximizing multipath routing throughput is a NP-complete problem due to the shared medium for electromagnetic wave transmission in wireless channel, inducing collision-free scheduling as part of the optimization problem. In this work, we first provide problem formulation that incorporates collision-free schedule, and then based on this formulation we design an algorithm with search pruning that jointly optimizes paths and transmission schedule. Though suboptimal, compared to the known optimal single path flow, we demonstrate that an efficient multipath routing scheme can increase the routing throughput by up to 100\% for simple WMNs.
\end{abstract}

\section{Introduction} \label{sect:Introduction}

A wireless mesh network (WMN) is a configuration of multihop self-organising nodes interconnected through wireless links. Because of the ease of deployment and cost-effectiveness, WMNs were initially developed for communications in scenarios such as battlefields and natural disasters, which requires rapid communication network deployment. It was later adopted for use in last mile network access to rural communities and wireless sensor networks (WSN).

Given a wide range of scenarios where WMN can be deployed, a lot of research contributions have been made to study and design efficient routing algorithm for WMN~\cite{Campista08}. Supporting high transmission bandwidth between a source and a destination in a WMN relies on finding high throughput path using routing algorithm. Finding the maximum multipath throughput in a WMN is a NP-complete problem under the constraint of collision-free scheduling~\cite{Jain03}. The broadcast nature of wireless transmission induces collision free transmission as part of the optimization problem. Majority of current routing algorithms for WMN are designed to find the single best path, often static, between a source and a destination. It is however intuitive to see that the performance boundary of the routing throughput can be exploited by optimizing additional routing paths.

Multipath routing algorithms increases the reliability of WMN by providing fault-tolerance due to node failure and has been shown to be useful in extending the lifetime of battery-constraint WSN and hence increase data flow by load-balancing the energy consumed for data flow on each of the paths~\cite{An11}. Flow transmission confidentiality in WMN can also be statistically increased by splitting the original encrypted information, and transmitting it along multiple paths between the source and the destination~\cite{Othman10}. This way, even if a malicious user has access to information from one of these path flows, the probability of the original message getting reconstituted would be relatively lower.

However despite the popularity of WMN over the last decade and the significance of multipath routing, results on the throughput bounds achievable by a wireless multipath routing algorithm has not been well studied. The purpose of this paper is to address this gap. In this paper we consider the problem of maximizing the routing throughput between any pair of nodes in an arbitrary WMN. We attempt to solve this problem by first building up a set of transmission constraints to guarantee collision free reception. We then establish useful search pruning axioms to significantly reduce the computation cost of finding paths. Next we use a greedy algorithm to spatially reuse `existing' time slots, and select only those additional routing paths which will increase the total routing throughput. We conduct simulation to test our algorithm and show that under the constraint of preestablished routing path flows, our algorithm can find throughput improving routing path, if one such path exists.

We catalog the findings of our work as follow. In Section~\ref{sect:Relatd} we first present a bibliography of related work on this topic. We then present problem formulation and network assumptions in Section~\ref{sect:Modelling}. We propose our algorithm in Section~\ref{sect:Algorithm} with an illustrating example. Simulation results to validate the performance of our algorithm is given in Section~\ref{sect:Simulation}. Finally we summarize key results of our findings in Section~\ref{sect:Conclusion}.

\section{Related Work} \label{sect:Relatd}
The first early works on maximum routing throughput for WMN was studied by Jain \emph{et al.}~\cite{Jain03} and Kodialam and Nandagopal~\cite{Kodialam03}. Jain \emph{et al.} showed that the maximum throughput problem in WMN can be reduced to the maximum independent set problem under the assumption of \textit{uniform} link capacities~\cite[Theorem 1]{Jain03}. It is easy to verify by constructing simple topology example that for \textit{multirate} WMN, transmission schedule on maximum independent set vertices need not necessarily lead to maximum routing throughput. Therefore heuristic algorithms for maximum independent set problem do not provide an efficient solution to maximize routing throughput in a multirate WMN.

Kodialam and Nandagopal developed algorithms to route maximum data between a source and a destination by jointly solving the routing and scheduling problem. However in their model, they only assume primary interference, and free the system from any secondary interference constraint. \emph{Primary interference} means that each node can communicate with at most one node during any time interval, and \emph{secondary interference} refers to the interference resulting from communications between different nodes. Secondary interference occurs when two or more simultaneous transmissions appear too close in space such that the receiver of a transmission is interfered by the sender of another simultaneous transmission. To assume that a wireless network is free of secondary interference is clearly impractical. Extension works of Kodialam and Nandagopal, such as the work of Chen \emph{et al.}~\cite{Chen06} to design a cross-layer jointly optimized congestion control, routing and scheduling algorithm for WMN, also considers only the primary interference model.

Where efforts have been made to calculate the maximum end-to-end routing throughput in a WMN under the constraint of both primary and secondary interference, such works have been limited to simple scenarios such as the chain-topology~\cite{Yoo07} or for specific routing techniques such as the opportunistic routing~\cite{Zeng08}. While there exists works~\cite{Krishnan08,Augusto11} to jointly optimize the routing and interference-free scheduling problem for an arbitrary WMN topology, such work optimizes the \emph{single best} path between each pair of the source-destination pairs. The works of Wan~\cite{Wan09} consider the problem of finding the maximum multiflow in a WMN with multiple pairs of source and destination, assuming each of these flow between a source and a destination follows one path, and therefore do not exploit the benefit of multiple paths between a source and a destination.

We compare the performance of our proposed algorithm with the medium time metric (MTM) routing scheme~\cite{Awerbuch06}. In this work the authors argue that majority of current on-demand and proactive routing protocols (see reference therein~\cite{Awerbuch06}) were designed for single-rate networks, and therefore have used the shortest path algorithm as a metric to select path. However, as they demonstrate, for a multirate network, a routing protocol which minimizes the total medium time to transmit data, optimally maximizes the end-to-end path throughput. Therefore MTM finds the optimal single routing path in a multirate WMN. We will therefore evaluate the throughput improvement of multipath routing over MTM optimal single path routing.

\section{The Model} \label{sect:Modelling}

\subsection{Network and Flow Model}

Nodes and connectivity links in network topology is represented by a directed connectivity graph $G=(V,L)$, where $V$ is the set of nodes, and $L$ is the set of directed links between these nodes. Cardinality of $V$ is given as $|V|$, and denotes the size of the network. The source and destination nodes are denoted by $R_s$ and $R_d$ respectively. Transmission link from $R_i$ to $R_j$, $1\leq i, j\leq |V|, i\neq j$, is represented by $l_{ij}$, where $l_{ij}\in L$. Let $c_{ij}$ be the link capacity on link $l_{ij}$ which is specified in bits per second ($bps$). We assume that bidirectional link capacity are equal, i.e. $c_{ij}=c_{ji}$. The set of neighbouring nodes connected to $R_i$ is given as $N(R_i)$. We assume that $N(R_i)\geq 2, \forall i\backslash\{R_s, R_d\}$, therefore no relay node is ``child-less''.

In our model we assume a fine-grained time allocation scheme, to efficiently reuse time slots over links of different capacities. The set of time slots is denoted as $T$ and these time slots may not necessarily be even in duration. Let there be $M$ elements in $T$, $M \geq 1$. Each of these elements is labeled as $t_k$, $1\leq k\leq M$, and its duration is given by $d(t_k)$ seconds.

For a given time slot $t_k$, the quantity $f_{ij,t_k}$ denotes useful unicast transmission information from node $R_i$ to $R_j$ for the entire duration of the time slot $t_k$. The word ``useful'' here has been used to distinguish transmission reception from opportunistic listening due to the broadcast nature of wireless transmission. The unit of $f_{ij,t_k}$ is bits, and it is bounded as $f_{ij,t_k} \leq c_{ij} \cdot d(t_k)$.

\subsection{Problem formulation}
Maximum multipath routing throughput optimization can be formulated as follow,
\begin{equation}\label{eq:eq1}
\text{maximize } \frac{\sum_{t_k\in T}\sum_{R_j\in N(R_d)}
f_{jd,t_k}}{\sum_{t_k\in T} d(t_k)},
\end{equation}
subject to, at time slot $t_k$,

\begin{equation}\label{eq:eq1.2}
f_{jm, t_k}=0 : f_{mn,t_k}>0, \; R_j\in N(R_m),
\end{equation}

\begin{equation}\label{eq:eq2}
f_{mn,t_k} = \sum_{R_i\in N(R_n)} f_{in,t_k} + \sum_{R_i\in N(R_n)} f_{ni,t_k} : \; f_{mn,t_k}>0,
\end{equation}

\begin{equation} \label{eq:eq4}
\sum_{t_k\in T}\sum_{R_j\in N(R_i)} f_{ji,t_k} = \sum_{t_k\in
T}\sum_{R_j\in N(R_i)} f_{ij,t_k}, \forall R_i \backslash \{R_s,
R_d\},
\end{equation}

\begin{equation} \label{eq:eq5}
\sum_{t_k\in T}\sum_{R_j\in N(R_s)} f_{sj,t_k} = \sum_{t_k\in
T}\sum_{R_j\in N(R_d)} f_{jd,t_k}.
\end{equation}

There are two parameters for optimization in Equation \eqref{eq:eq1}. Our first objective is to maximize the total data flow to the destination, and simultaneously minimize the total time duration required to achieve the first objective. By optimizing these two parameters jointly, we maximize the routing throughput.

Equation \eqref{eq:eq1.2} characterizes that if $R_m$ is transmitting to one of its neighbour $R_n$, then $R_m$ can not receive a transmission from any of its neighbour $R_j$. The first term of the Equation \eqref{eq:eq2} characterizes that when $R_n$ is receiving a transmission from $R_m$ then none of the other neighbouring nodes of $R_n$ can simultaneously transmit, and the second term characterizes that when receiving a transmission from $R_m$, $R_n$ cannot simultaneously transmit to any of its neighbor $R_i$. Therefore Equations \eqref{eq:eq1.2} and \eqref{eq:eq2} encapsulates both primary and secondary interference free reception.

The second set of constraints given in Equations \eqref{eq:eq4} and \eqref{eq:eq5} represents the conservation of flow at each node. Precisely, for \eqref{eq:eq4}, it states that for nodes in the graph $G$ except the source and the destination, a node cannot transmit an amount of useful information that exceeds what it has received. For \eqref{eq:eq5}, it states that the total amount of useful information transmitted by the source must be the same as the total amount of useful information received by the destination, which is the equivalent to the conservation of flow in network.

\subsection{Modeling Assumptions}\label{sec:modeling}
In our model, transmission is characterized as omnidirectional radio propagation. We assume that nodes' location are static and do not consider node mobility, which is applicable for mobile ad-hoc network (MANET). We characterize the secondary interference using the protocol interference model~\cite{Gupta00} where the transmission range equals the interference range. In a protocol interference model, a transmission is successful if the receiver lies within the transmission range of the sender, and concurrently does not lie within the interference range of any other simultaneously transmitting node.

In the network, only the source node $R_s$ can create the packet. We assume that the source node always have packets to transmit. Intuitively we give no credit for the destination node receiving duplicate packets. We assume that the nodes can perfectly schedule their transmission, using a perfect MAC protocol, which is consistent with assumptions used in previous seminal works~\cite{Gupta00,Jain03}. We assume multirate transmission in our model, as current transmission standards such as the IEEE 802.11a/b/g used in WMN support multirate transmission capability.

\section{Algorithm}\label{sect:Algorithm}
\newtheorem{theorem}{\textbf{Theorem}}
\newtheorem{corollary}{\textbf{Corollary}}
\newtheorem{axiom}{\textbf{Axiom}}

A path flow $p(f)\subseteq L$, is a tuple set of links forming an uninterrupted path from $R_s$ to $R_d$, where $f$ represents the path flow id, $1\leq f\leq g$, and $g$ is the total number of paths selected. The set $c_{ij}(f)$ is generated from the corresponding set $p(f)$ using a bijective function, where each element of $c_{ij}(f)$, $c_{ij}\in c_{ij}(f)$, represents the capacity of link $l_{ij}$, $l_{ij}\in p(f)$.

\begin{axiom}\label{axiom0}
\emph{The path flow value on path $p(f)$ is determined by
$min\{c_{ij}(f)\}$, i.e. the bottleneck link member of set $p(f)$.}
\end{axiom}

Based on Axiom \ref{axiom0}, our optimization problem \eqref{eq:eq1} can have a slightly modified description,
\begin{equation}\label{eq:condition1}
\text{maximize } \frac{\sum_{f=1}^{g}min\{c_{ij}(f)\}}{\sum_{t_k\in T} d(t_k)}.
\end{equation}
\noindent Given the set of preestablished path flows, an additional path flow $p(g+1)$ is added if it satisfies the following inequality condition,
\begin{equation}\label{eq:condition2}
\frac{\sum_{f=1}^{g+1}min\{c_{ij}(f)\}}{\sum_{t_k\in T}
d(t_k)+\delta}>\frac{\sum_{f=1}^{g}min\{c_{ij}(f)\}}{\sum_{t_k\in T} d(t_k)}.
\end{equation}

\noindent where $\delta$ is the additional time duration required to establish the $g+1^{th}$ path. Specifically inequality \eqref{eq:condition2} requires that an additional path be only added, if it can improve the overall routing throughput, given the set of preestablished path flows.

Therefore unlike the works of~\cite{Jain03,Chen06} which try to optimize all the possible path flows simultaneously by reducing the scheduling problem to the maximum independent problem, which is a NP-complete problem, our greedy-based algorithm attempts to find the maximum routing throughput problem by finding one path flow at a time, before searching for an additional path flow. Time slots are simultaneously adjusted during the process. This significantly reduce the computational cost of finding the maximum routing throughput.

\begin{axiom}\label{axiom1}
\emph{As the hop count of a path increases, the path throughput
either decreases or stays constant.}
\end{axiom}

When a hop is added to a path, intra-flow interference may intensify. Collision-free transmission is accommodated by allocating more time slots which adversely effect the path bandwidth.

\begin{axiom}\label{axiom2}
\emph{Link $l_{ji}$ can be deleted from $L$ if the following
condition is satisfied, $\exists t_k$, such that $f_{ij,t_k}>0$.}
\end{axiom}

If link $l_{ij}$ has been selected for transmission, then conversely transmission over link $l_{ji}$ would only constitute `returning' the data back to $R_i$, which is a redundant process, hence link $l_{ji}$ can be deleted from $L$, as transmission on link $l_{ji}$ is unwarranted. Axiom \ref{axiom1} and \ref{axiom2} serves as search pruning steps to reduce the computation cost of the algorithm.

\subsection{Fine-Grained Spatially Reused Time Allocation}

\begin{figure*}
\begin{center}
\includegraphics[width = 0.85\textwidth]{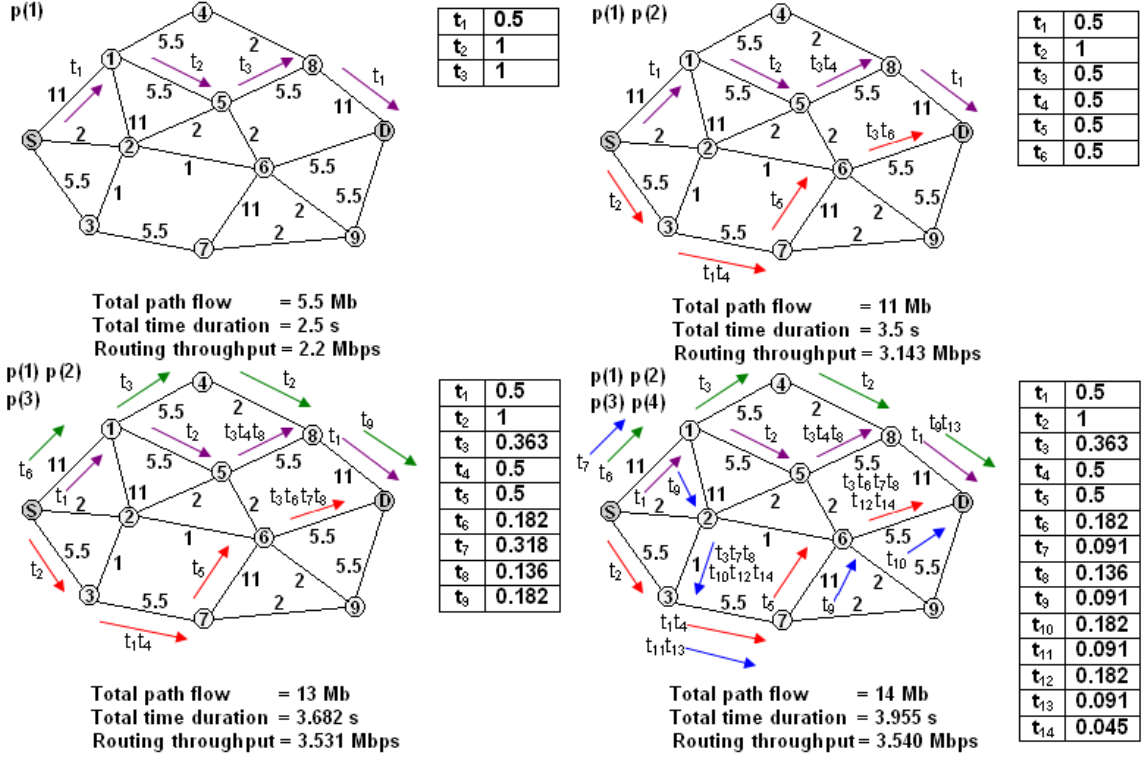}
\end{center}
\caption{An illustrating example to find multipath routing
throughput for $G=(11, 38)$, with $g=4$ path flows. All values have
been rounded to 3 decimal places. Source and destination nodes have
been marked with `S' and `D' respectively.} \label{fg:fig1}
\end{figure*}

\begin{table}
\caption{SpatialReuse($l_{mn}$)} \label{table:time0}
\begin{center}
\begin{tabular}{|l|}
\hline
// Time slots allocation on link $l_{mn}$.\\
\\
\ \ $t_k$.needed $\leftarrow$ $\frac{min\{c_{ij}(g+1)\}}{c_{mn}}$\\
\\
\ \  \textbf{For} $\forall t_i$, $t_i\in T$\\
\ \ \ \ \ \ \textbf{If} Equation \eqref{eq:eq2} is true for $f_{mn,t_i}$\\
\ \ \ \ \ \ \ \ \ \ $t_k$.available($t_i$)\\
\\
\ \ \ \ \ \ \ \ \ \ \textbf{If} ($|t_k$.available$|\geq|t_k$.needed$|$)\\
\ \ \ \ \ \ \ \ \ \ \ \ \ \ \textbf{Break}\\
\\
\ \ \textbf{If} $|t_k$.available$|$ = 0\\
\ \ \ \ \ \ $t_k$.create\\
\\
\ \ \textbf{Else If} ($|t_k$.available$|>|t_k$.needed$|$)\\
\ \ \ \ \ \ $t_k$.split($t_i$)\\
\ \ \ \ \ \ $t_k$.allocate($t_i$)\\
\\
\ \ \textbf{Else If} ($|t_k$.available$|\leq|t_k$.needed$|$)\\
\ \ \ \ \ \ $t_k$.allocate($t_i$)\\
\ \ \ \ \ \ $t_k$.create\\
\hline
\end{tabular}
\end{center}
\end{table}

We apply the concept of spatially reusing existing time slots. Once link $l_{mn}$ is selected by the algorithm for transmission, then from the existing time slots which had already been created, $\forall t_k$, we select those time slots would not result in any collision as per the constraint of Equation \eqref{eq:eq2} on link $l_{mn}$. 

When the total available time slots duration is more than the required time duration for transmission on link $l_{mn}$, then one of the existing time slot is split into two new time slots. Without loss of generality, those links which were previously allocated the time slot which had been splitted will now be allocated with both the splitted time slots. If no such time slot exist, or if the time time slots duration is insufficient for collision free transmission on $l_{mn}$, then new time slots are created accordingly. The pseudocode of the time duration allocation is given in Table \ref{table:time0}.

We illustrate this with an example, suppose transmission on $l_{mn}$ requires total time duration of 0.6$s$, however the available time slots $t_6$ and $t_4$ for collision free transmission on $l_{mn}$ have duration of 0.4$s$ and 0.3$s$ respectively. Therefore $t_4$ is split and assigned duration as $t_4=0.2$ and $t_7=0.1$. Time slots $t_6$ and $t_4$ is allocated to link $l_{mn}$. And those links on which $t_4$ was previously assigned is now assigned with time slots $t_4$ and $t_7$.

\subsection{Path Search}
We use the backtracking algorithm~\cite[pp. 230-239]{Skiena08} to find routing paths using Axioms~\ref{axiom0}-\ref{axiom2} as pruning techniques to minimize the computation cost. Unlike the exhaustive search algorithm which searches for a path from all possible permutation of paths from source to destination, in a backtracking algorithm, the search starting at the source node terminates several hops before reaching the destination if progressing any further on the partial path $p_p(f)$, $p_p(f)\subset p(f)$ does not increases the routing throughput based on Axiom~\ref{axiom1}. Consecutively for all paths whose subset is equal to $p_p(f)$ are ``blocked,'' and not searched by the backtracking algorithm.

So even though the backtracking algorithm has exponential complexity using the big O notation, its computation cost is a fraction of the computation cost of the exhaustive search algorithm. It can therefore be practically used for WMN, where the network size is in order of 100's of nodes. For larger networks, the Dijkstra's shortest path algorithm which finds the shortest distance (or maximum throughput for our case) from one node to all others in a weighted graph can used. This family of weighted graph traversal algorithms has time complexity of $O(|V|^2)$.

For either of the graph traversal algorithms proposed earlier, the algorithm runs Equation~\eqref{eq:eq1.2} and~\eqref{eq:eq2} to ensure collision free schedule, and then evaluates the tentative routing throughput by running SpatialReuse() algorithm given in Table~\ref{table:time0}. The path search algorithm finds one path at a time, before searching for additional path which will improve the total routing throughput.

\subsection{Illustrating example}
We illustrate the algorithm with the aid of a self-explanatory example shown in Figure~\ref{fg:fig1}, for which we have used the 802.11b transmission rates. The table on the right of each graph list all the time slots and its corresponding duration. The algorithm first finds path $p(1)$, resulting in a routing throughput of $2.2Mbps$. Time slot $t_1$ has been re-used for simultaneous transmissions on $l_{s1}$ and $l_{8d}$. The duration of time slots is calculated based on path flow and link capacity. For $p(1)$, $min\{c_{ij}(1)\}=5.5Mbps$, therefore transmitting $5.5Mb$ on link $l_{s1}$ and $l_{8d}$ requires duration of $0.5s$, hence $t_1=0.5s$. 

The algorithm then attempts to add an additional path $p(2)$. Additional of $p(2)$ doubles the total flow from $5.5Mb$ to $11Mb$, however by spatially re-using the time slots allocated for path $p(1)$, total time duration does not increase by the same factor as total flow, thus the routing throughput increases from $2.2Mbps$ to $3.143Mbps$. While adding path $p(2)$, time slot $t_3$ is split as $t_3=0.5$ and $t_4=0.5$. Continuing this way, for the final solution of four paths, collision-free transmissions schedule is designed to maximize routing throughput, and no further path can be added to increase the routing throughput.

\section{Simulation}\label{sect:Simulation}
\begin{figure}
\begin{center}
\includegraphics[width = 0.45\textwidth]{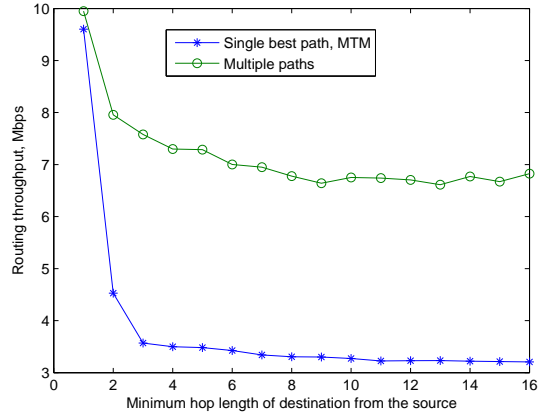}
\end{center}
\caption{Routing throughput with respect to the minimum hop length distance between the source and the destination, for $G=(100,320)$.} \label{fig:graph1}
\end{figure}

We construct a discrete-time simulator with implementation of our algorithm to find maximum routing throughput in a WMN under modeling assumptions listed in Section~\ref{sec:modeling} using the backtracking algorithm. For comparison, we implement MTM routing scheme which has been shown to optimize the single-best routing path in a multirate multihop WMN~\cite{Awerbuch06}. We construct a 100 nodes WMN, where the nodes have been randomly located. The link capacity of each direct connection is randomly and uniformly assigned with a value between 5 Mbps and 15Mbps in intervals of 1Mbps. The source and destination pair is randomly selected. A typical routing throughput comparison is shown in Figure~\ref{fig:graph1}. Average running time for multipath backtracking algorithm on graph $G=(100,320)$ was approximately 0.5 seconds on a 2GHz processor system.

As shown in Figure~\ref{fig:graph1}, a one-hop distance between the source and destination gives similar maximum routing throughput. This is because the distance is too short to take advantage of multiple paths. As the distance increases, while the single path routing performance drops significantly due to primary interference, multipath routing drops slightly as it is able to find more collision-free paths to offset this interference. The performance converges at three-hop distance as the flow constraint is still mainly determined by primary interference. Small fluctuation is seen in multipath routing throughput after three-hop distance due to inter-flow interference.

\section{Conclusion}\label{sect:Conclusion}
In this paper we formulated optimization problem for maximum routing throughput for multirate WMNs under collision free schedule. The principal contribution of our algorithmic solution is to demonstrate the effectiveness of spatially reusing time slots towards improving the total routing throughput of WMN. By using search pruning technique, we are able to run the backtracking algorithm on WMN with 100s of nodes, which is practical for most applications. Though the proposed algorithmic solution is suboptimal, through simulation result we have shown that it can improve the routing throughput by up to 100\% for a simple $G=(100,320)$ WMN.

\bibliographystyle{IEEEtran}
\bibliography{IEEEabrv,mainJ}

\end{document}